\documentclass[aip,jcp,groupedaddress,reprint]{revtex4-1}

\usepackage{hyperref}
\usepackage{amsmath,amssymb}
\usepackage{bm}
\usepackage{graphicx}
\usepackage{color}

\newcommand{\ddr}{\ensuremath{\mathrm{d}}}                                 % \ddr means "d droit"
\newcommand{\nablab}{\ensuremath{\bm{\nabla}}}                             % \nabla in bold
\newcommand{\deriv}[2]{\ensuremath{\frac{\mathrm{d} #1}{\mathrm{d} #2}}}   % \deriv : d .../d ...
\newcommand{\pderiv}[2]{\ensuremath{\frac{\partial #1 }{ \partial #2}}}    % \pderiv : \deriv with \partial
\begin{document}

%%% TITLE HEADER %%%
\title{Lack of an equation of state for the nonequilibrium chemical potential of gases of active particles in contact}

\date{\today}

\author{Jules \surname{Guioth}}
\email{jules.guioth@damtp.cam.ac.uk}
\affiliation{DAMTP, Centre for Mathematical Sciences, University of Cambridge, Wilberforce Road, Cambridge CB3 0WA, UK}
\affiliation{Univ.~Grenoble Alpes, CNRS, LIPhy, F-38000 Grenoble, France}
\author{Eric \surname{Bertin}}
\email{eric.bertin@univ-grenoble-alpes.fr}
\affiliation{Univ.~Grenoble Alpes, CNRS, LIPhy, F-38000 Grenoble, France}

\begin{abstract}
We discuss the notion of nonequilibrium chemical potential in gases of non-interacting active particles filling two compartments separated by a potential energy barrier. Different types of active particles are considered: run-and-tumble particles, active Brownian particles, and active Brownian particles with a stochastic reorientation along an external field.
After recalling some analytical results for run-and-rumble particles in one dimension,
we focus on the two-dimensional case and obtain a perturbative expression of the density profile in the limit of a fast reorientation dynamics, for the three models of active particles mentioned above.
Computing the chemical potentials of the non-equilibrium systems in contact from the knowledge of the stationary probability distribution of the whole system
---which agrees with a recently proposed general definition of the chemical potential in non-equilibrium systems in contact--- we generically find that the chemical potential lacks an equation of state, in the sense that it depends on the detailed shape of the potential energy barrier separating the compartments and not only on bulk properties, at odds with equilibrium.
This situation is reminiscent of the properties of the mechanical pressure in active systems. We also argue that the Maxwell relation is no longer valid and cannot be used to infer the nonequilibrium chemical potential from the knowledge of the mechanical pressure.
\end{abstract}


\maketitle 
%%% TITLE HEADER (END) %%%
%%%%%%%%%%%%%%%%%%%%%%%%%%%%%%%%%%%%%%%%%%%%%%%%%%%%%%%%%
%%%%%%%%%%%%%%%%%%%%%%%%%%%%%%%%%%%%%%%%%%%%%%%%%%%%%%%%%

%%% INTRODUCTION %%%
\section{Introduction}
\label{sec:intro}

Active matter systems like assemblies of macroscopic particles in which energy is continuously fed at the particle scale \cite{marchetti2013hydrodynamics}, are a useful benchmark to test nonequilibrium extensions of thermodynamic concepts \cite{Speck2016,takatori2015towards,solon2015nat,solon2018generalized-njp}.
These systems include both artificial particles like self-propelled colloids \cite{Palacci2010,Theurkauff2012,Bechinger2013,palacci2013living,Bartolo2013,Bechinger2016} or vibrated macroscopic particles \cite{Dauchot,Ramaswamy-exp,Kudrolli}, and biological systems like assemblies of bacteria for instance \cite{Staruss,Aranson}.
On the theoretical side, several simple models of non-interacting active particles have been introduced (see Ref.~\onlinecite{Bechinger-review,fodor2018statistical} for recent reviews). 
Two of these models have emerged as paradigmatic models of active particles,
namely the Run-and-Tumble Particle (RTP) model \cite{tailleur2008statistical,cates2013active,schnitzer1993theory} modeling the motion of bacteria like \emph{E. Coli}, and the Active Brownian Particle (ABP) \cite{cates2013active,Stenhammar2014} model describing active colloids for instance.

In recent years, it has been realized that even an apparently well-defined notion like mechanical pressure can exhibit an unexpected behavior, as it lacks an equation of state \cite{fily2017mechanical,solon2018generalized-njp,solon2015nat,solon2015prl,speck2016ideal,takatori2014swim,winkler2015virial,Joyeux2016}.
This means that unlike at equilibrium, the mechanical pressure exerted on a wall depends not only on the bulk density, but also on the detailed shape of the diverging potential energy profile defining the wall. This dependence results from a nonequilibrium density profile within the (soft) wall, and can be ultimately traced back to the lack of momentum conservation in the bulk \cite{fily2017mechanical}, which is a consequence of the self-propulsion mechanism.

At equilibrium, intensive parameters (like temperature, pressure, chemical potentials,...) are generally defined as derivatives of the free energy with respect to their associated extensive variable (resp. energy, volume, number of particles,...). However, out of equilibrium, the free energy is in general no longer defined and the natural thermodynamic definition can no longer be applied.
There are essentially two ways to circumvent this issue. On the one hand, one can apply operational definitions. For instance, pressure can always be considered as the average mechanical force per unit area on the surface of the container. On the other hand, one can try to generalize the thermodynamic definition which relies on the equalization of intensive parameters when two systems in contact are allowed to exchange a conserved quantity, e.g.,
energy for temperature\cite{Bertin04,Levine2007} or number of particules for chemical potentials\cite{Bertin06,Bertin07}.
As energy is generically not conserved in driven systems, there is no clear way to extend the validity of the equilibrium definition along this line.
In contrast, the number of particles is often conserved even for driven systems, offering a way to define a nonequilibrium chemical potential (whose operational definition in terms of work is less straightforward than for the pressure).
Based on a large deviation approach, a precise definition of such a nonequilibrium chemical potential has been proposed recently for generic driven systems in weak contact.\cite{Guioth2018,Bertin06,Bertin07}. This well-grounded definition notably helps clarifying some of the results reported in the context of lattice gas models in contact.\cite{Guioth2018,Seifert11,Dickman14} The main result is that the nonequilibrium chemical potential defined in this way equalizes by construction between the two systems in contact, but generically depends on the details of the contact dynamics.\cite{Guioth2018}

Here, we wish to explore the consequences of this definition of a nonequilibrium chemical potential in the more realistic context of systems of active particles, for which experimental realizations are available\cite{Palacci2010,Theurkauff2012,Bechinger2013,palacci2013living,Bartolo2013,Bechinger2016,Dauchot,Ramaswamy-exp,Kudrolli} (see also the very recent work [\onlinecite{Pradhan2019}] for a related attempt in the Vicsek model).
In the spirit of the seminal work of Sasa and Tasaki on the generalization of thermodynamics to nonequilibrium systems,\cite{sasa2006steady} a weak contact can be realized between two systems by connecting them through a high potential energy barrier.
Such a setting can be easily implemented for active particles.
With this goal in mind, we consider three distinct models of non-interactive active particles. The first two models are standard models, namely the Run-and-Tumble Particle (RTP) model, and the Active Brownian Particle (ABP) model
\cite{cates2013active}.
In both cases, particles have an overdamped dynamics with a velocity characterized, in two dimensions, by an angle $\theta(t)$. 
In the presence of an external potential $U({\bf r})$, the position ${\bf r}(t)$ of a particle evolves according to
\begin{equation} \label{eq:def:dynamics}
\dot{\bf r} = v_0(\mathbf{r})\, {\bf e}(\theta) - \eta \nabla U \,,
\end{equation}
with ${\bf e}(\theta)$ the unit vector in the direction $\theta$,
$v_0(\mathbf{r})$ the (possibly space-dependent) speed of the particles, and $\eta$ is the translational mobility coefficient.
The dynamics of the angle $\theta(t)$ depends on the model.
For the RTP dynamics, a new value of the angle $\theta$ is drawn from a uniform distribution over $[0,2\pi)$ with a probability rate $\alpha$.
The run-and-tumble moded can also be defined in one dimension, in which case the angle $\theta$ can only take the values $0$ and $\pi$.
In the case of ABPs, the angle $\theta$ continuously diffuses,
\begin{equation} \label{eq:angular:diff}
\dot{\theta} = \xi(t)
\end{equation}
where $\xi(t)$ is a white noise satisfying $\langle \xi(t)\rangle =0$ and
\begin{equation}
\langle \xi(t)\xi(t')\rangle = 2D_{\rm r} \, \delta(t-t') \,,
\end{equation}
with $D_{\rm r}$ the angular diffusion coefficient.
As the angle $\theta$ varies continuously, no one-dimensional version of the ABP model can be defined.

As a third model, we generalize the ABP dynamics by adding to the diffusive dynamics of $\theta$ given by Eq.~(\ref{eq:angular:diff}) a stochastic jump dynamics that tends to align the velocity along an externally given direction. This model is inspired by experiments on microalgae \emph{Chlamydomonas Reinardtii} \cite{Rafai2016}, that randomly reorient their swimming direction towards the direction $\theta_0$ of a light source, with some noise in the reorientation process. The model is thus defined as an ABP model of angular diffusion coefficient $D_{\rm r}$, with a random reorientation, at rate $\lambda$, to a new direction $\theta'$ drawn from a distribution $\psi(\theta-\theta_0)$.
%For the sake of simplicity, we assume that the distribution $\psi$ is symmetric, i.e., $\psi(-\theta)=\psi(\theta)$.

The definition of a nonequilibrium chemical potential proposed in Refs.~\onlinecite{Bertin06,Bertin07,Guioth2018} relies on a large deviation formalism, that we briefly recall here.
Splitting a given system into two subsystems A and B such that the total number of particles $N_A+N_B=N$ is conserved (while particles can be exchanged between A and B), one determines the distribution $P(N_A|N)$ under the asymptotic form, valid for a large volume $V=V_A+V_B$,
\begin{equation}
P(N_A|N) \sim e^{-V \mathcal{I}(\rho_A,\rho_B)}
\end{equation}
where $\rho_A=N_A/V_A$ and $\rho_B=N_B/V_B$ are constrained by the conservation law $\gamma_A \rho_A+\gamma_B\rho_B=\overline{\rho}$, with $\gamma_k=V_k/V$ and $\overline{\rho}$ the overall density.
The function $\mathcal{I}(\rho_A,\rho_B)$ is called a large deviation function.
In the weak contact limit, when the rate of particle exchange is very small (for instance due to a high energy barrier separating the two systems), and when a single particle is exchanged at a time, it has been shown\cite{Guioth2018} that the large deviation function $\mathcal{I}(\rho_A,\rho_B)$ is additive,
\begin{equation} \label{eq:additivity}
\mathcal{I}(\rho_A,\rho_B) = \gamma_A I_A(\rho_A) + \gamma_B I_B(\rho_B),
\end{equation}
provided that the contact dynamics satisfies a factorization property with respect to compartments A and B.
When the additivity property of the large deviation function holds,
nonequilibrium chemical potentials can be defined from the relation
\begin{equation} \label{eq:def:chem:pot}
I_A'(\rho_A) - I_B'(\rho_B) = \mu_A^{\rm cont} - \mu_B^{\rm cont}
\end{equation}
for systems A and B in contact (the prime denotes a derivative).
The goal of this paper is thus to determine the large deviation function $\mathcal{I}(\rho_A,\rho_B)$ in the different models of active particles mentioned above, and to deduce from it the chemical potentials $\mu_k(\rho_k)$, $k=A$, $B$.
Note that in the approach presented here, the evaluation of $\mathcal{I}(\rho_A,\rho_B)$ requires first to determine the stationary density profile $\rho(x)$.
As a result, the obtained chemical potentials are not used to predict the stationary densities in systems A and B in contact, as these stationary densities are already known from the density profile (yet, equalizing the chemical potentials yields back the stationary densities).
The focus of our study is rather to determine whether the nonequilibrium chemical potentials that we have defined obey an equation of state, in the case of the simplest systems of active particles.

On the technical side, the lack of time-reversibility (or detailed balance) in the dynamics ---which appears in the presence of a non-uniform potential as we will see henceforth---  makes it difficult to compute the stationary density profile, even for independent particles. However, perturbative computations may be feasible in certain limits of the parameters. This is the case in particular when the reorientation time scale is very small compared to the translational time scales as we will see in Sect.~\ref{sec-2D} for the three models defined above.
Before presenting this generic perturbative method, we recall in Sect.~\ref{sec-1D} the exact results obtained for the one-dimensional run-and-tumble model, and discuss how the potential energy barrier unveils the otherwise hidden irreversibility of the dynamics \cite{fodor2016far,nardini2017entropy}.

%%%%%%%%%%%%%%%%%%%%%%%%%%%%%%%%%%%%%%%%%%%%%%%%%%%%%%%%%%%%%%%%
%%%%%%%%%%%%%%%%%%%%%%%%%%%%%%%%%%%%%%%%%%%%%%%%%%%%%%%%%%%%%%%%

\section{Run-and-Tumble particles in 1D}
\label{sec-1D}

In one dimension, there are only two possible directions of motion, namely left or right. Calling $P_{L}(x,t)$ and $P_{R}(x,t)$ the probability density distributions associated with a velocity oriented to the left or to the right respectively, the master equation reads

\begin{subequations}
  \label{eq:master_eq_RTP_1D}
  \begin{align}
    \pderiv{P_{R}}{t}(x,t) = & - \pderiv{}{x}\left[ \left( v_{0}(x) - \eta U'(x) \right) P_{R}(x,t) \right] \label{eq:master_eq_RTP_1D-R} \\
    & - \alpha\left( P_{R}(x,t) - P_{L}(x,t) \right)  \notag \\
    \pderiv{P_{L}}{t}(x,t) = & - \pderiv{}{x}\left[ \left( -v_{0}(x) - \eta U'(x) \right) P_{L}(x,t)\right] \label{eq:master_eq_RTP_1D-L} \\
                         &  + \alpha\left(P_{R}(x,t) - P_{L}(x,t) \right)  \; . \notag
  \end{align}
\end{subequations}
Defining $P(x,t)=P_{R}(x,t)+P_{L}(x,t)$ and $M(x,t)=P_{R}(x,t)-P_{L}(x,t)$, equations \eqref{eq:master_eq_RTP_1D} become
\begin{subequations}
\label{eq:master_eq_RTP_1D_dens_and_diff}
\begin{align}
   \pderiv{P}{t}(x,t)  & =  - \pderiv{J}{x}(x,t) \label{eq:master_eq_RTP_1D_dens_and_diff-P} \\
   \pderiv{M}{t}(x,t)  & =   - \pderiv{}{x}\left[ v_{0}(x)P(x,t) - \eta U'(x)M(x,t)\right] \label{eq:master_eq_RTP_1D_dens_and_diff-D} \\
  & \hspace{1.1em} - 2\alpha M(x,t) \; ,  \notag
\end{align}
\end{subequations}
with $ J(x,t) = v_{0}(x)M(x,t) - \eta U'(x)$.

Now, let us formally split our system into two parts A and B that are separated by a high barrier $U(x)$.
In this one-dimensional context, we define compartment A as the interval
$(-L_A,0)$ and compartment B as the interval $(0,L_B)$.
 One can also assume that $v_{0}(x)=v_{A}$ if $-L_A<x<0$ and $v_{0}(x)=v_{B}$
if $0<x<L_B$, i.e., that the velocities in A and B can be different. This situation may be particularly relevant for light-controlled self-catalytic propulsion as described for instance in Ref.~\onlinecite{Bechinger2013,palacci2013living,stenhammar2016light,Bechinger2016}, where the light intensity could differ in compartments A and B, leading to different self-propulsion speeds (note however, that ABP are a better description of active colloids that RTP\cite{cates2013active}).

In the bulk of each system, where the potential $U(x)$ is flat, the current reads $J(x,t)=v_{A,B} M(x,t)$. Since the boundaries are closed, the current is vanishing at the boundaries. The question we ask is how the two densities $\rho_{A}$ and $\rho_{B}$ in the bulk of each system are related to each other when particles are allowed to cross the potential barrier $U(x)$ centered at $x=0$.

To address this question, we look for a solution in the presence of the potential barrier. 
For the self-propulsion to be strong enough so that a particle can overcome the barrier, the condition $v_{0}(x)>\eta |U'(x)|$ has to be fulfilled for all $x$.
If this is the case, it has been shown \cite{tailleur2008statistical} that a steady-state solution of equations \eqref{eq:master_eq_RTP_1D_dens_and_diff} reads, on an interval where $v_{0}(x)$ takes a constant value $v_0$,
\begin{align}
  \label{eq:steady_state_sol_RTP_1D}
  & P(x) \propto \frac{1}{1-\left( \tfrac{\eta U'(x)}{v_{0}} \right)^{2}}  \\
  & \qquad \hspace{1.2em} \times \exp \left[ - \int^{x}\!\! \ddr q \, \left(\frac{\eta \alpha}{v_{0}^{2}} \right) \frac{U'(q)}{1-\left(\tfrac{\eta U'(q)}{v_{0}(q)} \right)^{2}} \right] \, , \notag
\end{align}
with $P(x)=0$ if $v_{0}(x)<\eta |U'(x)|$. One can see that this solution is associated with a vanishing current $J(x)$. Note that an accumulation of particles occurs around the potential energy barrier (see, e.g., Ref.~\onlinecite{fodor2018statistical}), which already suggests that out-of-equilibrium effects are present near the separating barrier.

In order to compute the stationary distribution of density in $A$ and $B$, one can proceed in two ways. The first one is straightforward: as one knows the exact stationary distribution for the whole system, one can compute exactly the probability density of a configuration $\{X_{i}\}_{i=1}^{N}$ and then find the probability to have $N_{A}$ particles in $A$. In a word, this way is the one someone would apply in an equilibrium situation where the Gibbs-Maxwell-Boltzmann distribution is known. The second one proceeds in a similar fashion as Ref.~\onlinecite{Guioth2018} and focuses directly on the exchange dynamics at the contact area.

\subsection{Evaluation of the chemical potentials from the density profile}

In the case described above of two compartments with speed $v_A$ and $v_B$, the stationary density field $\rho(x)=NP(x)$ reads for $x\in \Lambda_{k}$ ($k=A,\,B$), using Eq.~(\ref{eq:steady_state_sol_RTP_1D})
\begin{align}
  \label{eq:stationary_density_RTP_1D}
  \rho(x) & =  \frac{\rho_{k}^{\ast}}{1-\left( \tfrac{\eta U'(x)}{v_{k}} \right)^{2}} \\
  & \hspace{1em} \times \exp\left[-\int_{x_{k}^{\ast}}^{x}\!\!\ddr q\,  \frac{\eta \alpha}{v_{k}^{2}} \frac{U'(q)}{1-\left( \tfrac{\eta U'(q)}{v_{k}} \right)^{2}} \right] \, , \notag
    % \intertext{and for  $x \in \Lambda_{B}$}
    % \rho(x) & =  \frac{\rho_{B}^{\ast}}{1-\left( \tfrac{\eta^{\rm tr}U'(x)}{v_{B}}\right)^{2}}\exp\left[-\int_{x_{B}^{\ast}}^{x}\!\!\ddr q\, \frac{\eta^{\rm tr}\alpha}{v_{B}^{2}} \frac{U'(q)}{1-\left( \tfrac{\eta^{\rm tr} U'(q)}{v_{B}} \right)^{2}}\right] \; . \label{eq:stationary_density_RTP_1D-B}
\end{align}
with $x_{k}^{\ast}$ a point in the bulk of $k$ where the potential barrier $U(x)=0$, and $\rho_{k}^{\ast}$, the densities in the bulks to be determined. The current being continuous at contact between $A$ and $B$, the constants $\rho_{A}^{\ast}$ and $\rho_{B}^{\ast}$ are not independent from each other but are related by the equality of the stationary fluxes of particles going from $A$ to $B$ and from $B$ to $A$, yielding
\begin{equation}
  \label{eq:equalization_current_RTP_1D}
  v_{A}\rho_{A}^{\ast} e^{-\Delta Q_{A}} = v_{B}\rho_{B}^{\ast} e^{-\Delta Q_{B}}
\end{equation}
with
\begin{equation}
  \label{eq:delta_Q-correction_RTP_1D}
  \Delta Q_{k} = \int_{x_{k}^{\ast}}^{0} \!\!\ddr q \, \frac{\alpha \eta}{v_{k}^{2}} \frac{U'(q)}{1 - (\eta U'(q)/v_{k})^{2}}  \; . 
\end{equation}
At the thermodynamic limit where $L_{A}$ and $L_{B}$ tend to infinity, the normalization condition of $\rho(x)$ reads
\begin{equation}
  \label{eq:normalization_condition_RTP_1D}
  N = \int_{-L_A}^{L_B} \ddr x \rho(x) = L_{A}\rho_{A}^{\ast} + L_{B}\rho_{B}^{\ast}
\end{equation}
since the non-uniformity caused by the external potential $U(x)$ stays localized around the contact area. Equations \eqref{eq:equalization_current_RTP_1D} and \eqref{eq:normalization_condition_RTP_1D} of course completely determine $\rho_{A}^{\ast}$ and $\rho_{B}^{\ast}$ but one would like to see how chemical potentials of systems in contact can be defined in this simple case. 

Knowing the density field $\rho(x)$, one can now calculate the probability $P(N_{A}|N)$ to observe $N_{A}$ particles in compartment A (and thus $N_{B}=N-N_{A}$ particles in B). Particles being independent, the corresponding probability distribution is simply a binomial
\begin{equation}
  \label{eq:probability_number_particle_A_RTP_1D}
  P(N_{A}|N) = \binom{N}{N_{A}}\, p_A^{N_A} \, p_B^{N-N_A}
\end{equation}
where $p_A$ and $p_B$ are the probabilities for a given particle to be in compartment $A$ or $B$ respectively.
These probabilities are simply expressed in terms of the density field $\rho(x)$ as
\begin{align}
p_A &= \frac{1}{N} \int_{-L_A}^{0}\!\! \ddr x \, \rho(x) \,, \\
p_B &= \frac{1}{N} \int_{0}^{L_B}\!\! \ddr x \, \rho(x) \,.
\end{align}
In the thermodynamic limit, the distribution $P(N_{A}|N)$ takes the large deviation form
\begin{equation} \label{eq:large_dev_proba_density_A_RTP_1D}
 P(\rho_{A} | \bar{\rho} ) \sim e^{-L\mathcal{I}(\rho_{A},\rho_{B})}
\end{equation}
with $L=L_{A}+L_{B}$, and a large deviation function
\begin{equation} \label{eq:large_dev_proba_density_A_RTP_1D_I}
\mathcal{I}(\rho_{A},\rho_{B}) = \gamma_{A}\rho_{A}\ln\frac{\rho_{A}}{\rho_{A}^{\ast}} + \gamma_{B}\rho_{B} \ln\frac{\rho_{B}}{\rho_{B}^{\ast}}
\end{equation}
where $\rho_A$ and $\rho_B$ are constrained by
\begin{equation} \label{eq:density:constraint}
\gamma_{A}\rho_{A}+\gamma_{B}\rho_{B} = \bar{\rho} \,,
\end{equation}
with $\gamma_k=L_k/L$.
%
%To derive Eq.~(\ref{eq:large_dev_proba_density_A_RTP_1D_I}), we have used the fact that
%\begin{equation}
%\int_{-L_A}^{0}\!\! \ddr x \, \rho(x) \to \rho_A^{\ast} \,, \quad
%\int_{0}^{\rho_B^{\ast}}\!\! \ddr x \, \rho(x) \to \rho_B^{\ast}
%\end{equation}
%in the thermodynamic limit $L_A$, $L_B\to \infty$.
%
Using Eq.~(\ref{eq:equalization_current_RTP_1D}), the large deviation function $\mathcal{I}(\rho_{A},\rho_{B})$ can be rewritten in the additive form
Eq.~(\ref{eq:additivity}), with $I_k(\rho_k)$ given by
\begin{equation}
I_k(\rho_k) = \rho_{k}\ln\frac{\rho_{k}}{\rho_{k}^{\ast}} \qquad (k=A,\,B).
\end{equation}
Taking the derivative of $\mathcal{I}(\rho_{A},\rho_{B})$ with respect to $\rho_A$ under the contraint (\ref{eq:density:constraint}), one obtains
\begin{equation}
\frac{1}{\gamma_A}\frac{d\mathcal{I}}{d\rho_A} = I_A'(\rho_A)-I_B'(\rho_B) = \ln\frac{\rho_A}{\rho_B}
- \ln \frac{\rho_A^{\ast}}{\rho_B^{\ast}} \,.
\end{equation}
Using Eq.~(\ref{eq:equalization_current_RTP_1D}) to reexpress the ratio $\rho_A^{\ast}/\rho_B^{\ast}$, one finds
\begin{equation}
I_A'(\rho_A)-I_B'(\rho_B) = \ln\frac{\rho_A v_A e^{-\Delta Q_A}}{\rho_B v_B e^{-\Delta Q_B}} \,.
\end{equation}
Since one has by definition
\begin{equation}
I_A'(\rho_A)-I_B'(\rho_B) = \mu_{A}^{\mathrm{cont}}(\rho_{A}) - \mu_{B}^{\mathrm{cont}}(\rho_{B}) \,,
\end{equation}
the chemical potential $\mu_{k}^{\mathrm{cont}}(\rho_{k})$ can be expressed as
\begin{equation}
  \label{eq:mu_cont_RTP_1D}
  \mu_{k}^{\mathrm{cont}}(\rho_{k}) = \ln \frac{\rho_{k}v_{k}}{\alpha}
-\Delta Q_{k} \,.
\end{equation}
In addition, the stationary densities $\rho_k^*$ obey
\begin{equation}
  \label{eq:derivative_chempot_RTP_1D}
   \mu_{A}^{\mathrm{cont}}(\rho_{A}^{\ast}) = \mu_{B}^{\mathrm{cont}}(\rho_{B}^{\ast})
\end{equation}
thanks to the property $d\mathcal{I}/d\rho_A(\rho_A^{\ast},\rho_B^{\ast})=0$.

The parameter $\alpha$ as been introduced in Eq.~(\ref{eq:mu_cont_RTP_1D}) as a typical rate (inverse time scale) to get a dimensionless argument in the logarithm; yet, the choice of this inverse time scale is arbitrary.
One notices that beyond the expected ``perfect gas'' contribution $\ln \rho_{k}$ in Eq.~(\ref{eq:mu_cont_RTP_1D}), a dependence on the particle speed appears, and more importantly a non-trivial dependence on the contact appears in the expression of a chemical potential at contact, through the term $\Delta Q_{k}$. Even when the velocities on each side are equals ($v_{A}=v_{B}$), $\Delta Q_{A}$ and $\Delta Q_{B}$ (see equation \eqref{eq:delta_Q-correction_RTP_1D}) are different as soon as the potential barrier $U(x)$ is asymmetric.
As a result, the chemical potential does not satisfy an equation of state, in the sense that it does not depend only on the bulk density, but also on the details of the contact dynamics.

\subsection{Computation based on the Poisson dynamics over the particle number}

As an alternative determination of the chemical potential, we would like to compute the large deviation function $\mathcal{I}(\rho_A,\rho_B)$ for the one-dimensional run-and-tumble particle model, by directly using a model of the exchange dynamics. The independence of the particles implies that the number of particles $N_{A}$ in system $A$ obeys a Poisson process. The associated master equation reads
\begin{align}
  \label{eq:master_eq_RTP_1D_bis}
  &\deriv{P}{t}(\rho_{A}|\bar{\rho})  \\
  & \hspace{2em} = v_{A} \rho(0^{-}) \left( P(\rho_{A}+\tfrac{1}{V_{A}}|\bar{\rho}) - P(\rho_{A}|\bar{\rho}) \right)  \notag \\
  & \hspace{3em} + v_{B}\rho(0^{+}) \left( P(\rho_{A}-\tfrac{1}{V_{A}}|\bar{\rho}) - P(\rho_{A}|\bar{\rho}) \right) \; , \notag
\end{align}
with
\begin{equation}
  \label{eq:rho_0-_0+}
  \rho(0^{-})  =  \rho_{A} e^{-\Delta Q_{A}} \quad \text{and} \quad \rho(0^{+}) = \rho_{B} e^{-\Delta Q_{B}} \; ,
\end{equation}
$\rho_{k}$ being the actual densities in the bulk of system $k$ such that $\gamma_{A}\rho_{A}+\gamma_{B}\rho_{B} = \bar{\rho}$, and $\Delta Q_{k}$ being defined in equation \eqref{eq:delta_Q-correction_RTP_1D}.
Note that $v_{k} \rho_{k}e^{-\Delta Q_{k}}$ is nothing but the average flux of particles (and also the transition rates here for this Poisson process) directed from $k\to k'\neq k$, $k,k' = A, B$.

Using the large deviation form (\ref{eq:large_dev_proba_density_A_RTP_1D}) in Eq.~(\ref{eq:master_eq_RTP_1D_bis}) yields an equation on the derivative of the large deviation function associated with $P(\rho_{A}|\bar{\rho})$.
Solving this equation, one finds
\begin{equation}
  \label{eq:HJ_sol_RTP_1D}
\frac{1}{\gamma_A}\frac{d\mathcal{I}}{d\rho_A} \equiv \mathcal{I}'(\rho_{A},\rho_B) = \ln \frac{\rho_{A} v_A e^{-\Delta Q_{A}}}{\rho_{B} v_B e^{-\Delta Q_{B}}}\,.
%=  \mu_{A}^{\mathrm{cont}} - \mu_{B}^{\mathrm{cont}}\;.
\end{equation}
From the relation $ \mathcal{I}'(\rho_{A},\rho_B) = \mu_{A}^{\mathrm{cont}} - \mu_{B}^{\mathrm{cont}}$, we thus recover, up to the introduction of the (arbitrary) normalizing time scale $\alpha$, the chemical potentials $\mu_{k}^{\mathrm{cont}}$ introduced in Eq.~\eqref{eq:mu_cont_RTP_1D}.

As emphasized in Ref.~\onlinecite{Guioth2018}, we note that the additivity of the large deviation function is a consequence of the factorization of the macroscopic transition rates (which are here of the so-called Sasa-Tasaki type\cite{sasa2006steady}) as well as macroscopic detailed balance, which results from the exchange of a single particle at a time between compartments A and B.

\subsection{Entropy production of 1D RTPs}

Entropy production in systems of active particles is a natural thermodynamic quantity to look at, because active particles dissipate energy and are thus expected not to obey time-reversal symmetry\cite{fodor2016far,nardini2017entropy,Seifert-entropyprod}.
With this idea in mind, it is interesting to note that from the point of view of trajectories, the run-and-tumble dynamics in the absence of external potential is time-reversible (however, the underlying processes that generate the self-propulsion are irreversible and do generate entropy production \cite{Seifert-entropyprod}, but they are not described explicitly in the run-and-tumble model).
One may thus wonder why some nonequilibrium effects would come into play. As seen above, nonequilibrium effects occur around the energy barrier, as the shape of the energy barrier matters for the definition of the chemical potentials of the systems in contact.
One thus expects that the presence of an external potential may break time-reversibility, and should thus lead to a non-zero entropy production.
Let us then compare trajectories with their time-reversal counterpart to examine the non-equilibrium character of this model.
%(the study of time-reversal symmetry in self-propelled particles and active matter has attracted recently a lot of interest, see for instance \cite{nardini2017entropy,fodor2016far}).
For one particle, a trajectory of length $T$ corresponds to the position $x(t)$ as well as the direction of its velocity $\pm v_{0}$, or equivalently the times of velocity reversals $\{t_{k}\}_{k=1}^{\mathcal{N}_{j}}$, over the interval $[0,T]$, $\mathcal{N}_{j}$ being the number of jumps during this interval. The position $x(t)$ thus obeys the equation of motion
  \begin{equation}
    \label{eq:eq_motion_RTP_1D}
    \dot{x}(t) = - \eta  U'(x(t)) + s(t)v_{0} \, ,
  \end{equation}
  with 
  \begin{equation*}
    s(t) = s_{k} = \pm 1 \qquad \text{for } t_{k}\leqslant t < t_{k+1} \, .
  \end{equation*}
  Calling $x^{\mathrm{R}}(t)=x(T-t)$ the time-reversed trajectory, it necessarily obeys the equation
  \begin{align}
    \label{eq:time_reversed_eq_motion_RTP_1D}
    \dot{x}^{\mathrm{R}}(t) = &- \eta U'(x^{\mathrm{R}}(t)) \\
   & + \underbrace{2\eta U'(x^{\mathrm{R}}(t)) - s(T-t)v_{0}}_{\notin \{+v_{0},-v_{0} \}} \; . \notag
  \end{align}
The question is to know whether this time-reversed trajectory obeys the same stochastic dynamics as the original one.
Clearly, the last term cannot be realised by the natural noise present in the system since the latter is only bi-valued, unless the external force vanishes ($U'(x)=0$). Thus, the presence of an external potential breaks the time-reversal symmetry and the entropy production gets infinite, because the probability to observe the time-reversed trajectory is simply zero. 
This infinite value of the entropy production is an artifact of the absence of thermal diffusion. In principle, a small thermal diffusive motion would come on top of the ballistic motion described by Eq.~(\ref{eq:def:dynamics}). Taking into account this thermal contribution restores a finite entropy production.
%Note also that the entropy production is not related here to any global current flowing in the system, but is necessary to maintain the non-equilibrium state. 

%%%%%%%%%%%%%%%%%%%%%%%%%%%%%%%%%%%%%%%%%%%%%%%%%%%%%%%%%%%%%%%%
%%%%%%%%%%%%%%%%%%%%%%%%%%%%%%%%%%%%%%%%%%%%%%%%%%%%%%%%%%%%%%%%

\section{Active dynamics in 2D: perturbative evaluation}
\label{sec-2D}

As mentioned in the introduction of Sec.~\ref{sec-1D}, very few situations even for independent particles allow explicit exact calculations. To see if the effect of the potential described before is not specific to the one-dimensional run-and-tumble dynamics, we now study several two-dimensional models of active particles, namely the two-dimensional run-and-tumble particle (RTP) model, the active Brownian particle (ABP) model, and the ABP with an external field.
We generically call $\Lambda_A$ and $\Lambda_B$ the regions of space defining the compartments A and B.

\subsection{Run-and-Tumble particles in 2D}

In two dimensions, relevant variables are the position $\mathbf{r}=x\bm{e}_{x} + y\bm{e}_{y}$ and the angle $\theta\in[0,2\pi)$. The master equation which describes the evolution of the probability density function $P(\mathbf{r},\theta,t)$ reads
\begin{align}
  \label{eq:master_eq_RTP_2D}
  & \pderiv{P}{t}(\mathbf{r}, \theta,t) \\
  &  \hspace{1em} = - \nablab \cdot \left[ (v_{0}(\mathbf{r}) \bm{e}(\theta) - \eta \nablab U(\bm{r}) ) P(\bm{r},\theta) \right] \notag \\
  & \hspace{2.1em}  + \frac{\alpha}{2\pi}\int_{0}^{2\pi} \!\! \ddr \tilde{\theta} P(\bm{r},\tilde{\theta}) - \alpha P(\bm{r},\theta). \notag
\end{align}
Finding the stationary solution of Eq.~(\ref{eq:master_eq_RTP_2D}) is difficult.
However, one expects an equilibrium-like behaviour in the limit where an infinite tumbling rate would lead to a translational diffusive-like behaviour (if the velocity $v_{0}$ scales in the proper way).
For the model to remain well defined, the effective diffusion coefficient $D\sim v_{0}^{2}/\alpha$ should remain finite when $\alpha\to\infty$. Our goal is then to compute the first order correction in $\alpha$ to this equilibrium-like limit. For the sake of simplicity and in order to see only the genuine out-of-equilibrium effect of an asymmetry of the potentiel barrier, we will assume here that $v_{0}(\bm{r})$ is uniform all along the whole system, $v_{0}(\bm{r})=v_0$.

To perform this study, we introduce several quantities and notations
(see Ref.~\onlinecite{peshkov2014boltzmann} for a review of these techniques).
First, we define the angular Fourier transform
\begin{align}
  \label{eq:def:Fourier_transform_2D}
  P(\mathbf{r},\theta,t) & = \frac{1}{2\pi}\sum_{k=-\infty}^{\infty} f_{k}(\mathbf{r},t)\, e^{-ik\theta} \\
  f_{k}(\mathbf{r},t) & = \int_{0}^{2\pi} \!\! \ddr \theta \, P(\mathbf{r},\theta,t)\, e^{ik\theta}, \notag
\end{align}
with $f_k^* = f_{-k}$, $z^*$ being the complex conjugate of $z\in\mathbb{C}$.
With these notations, one gets 
\begin{equation}
\rho(\mathbf{r},t) = \int_{0}^{2\pi}\ddr\theta P(\mathbf{r},\theta) = f_{0}(\mathbf{r})\,.
\end{equation}
Projecting the master equation \eqref{eq:master_eq_RTP_2D} onto the Fourier basis we have just introduced, gives a hierarchy of equations on $f_{k}$:
\begin{align}
  \label{eq:hierarchy_eqs_Fourier_RTP_2D}
  & \pderiv{f_{k}}{t} = - \frac{v_{0}}{2} \left(\triangledown f_{k-1} + \triangledown^* f_{k+1} \right) + \eta \nablab \cdot \left(f_{k} \nablab U \right) \notag \\
 & \hspace{2.2em} - \alpha f_{k} \left(1- \delta_{k,0} \right) ,
\end{align}
where one has introduced the complex derivatives
\begin{equation}
\triangledown = \partial_{x} + i\partial_{y} \, , \quad
\triangledown^* = \partial_{x} - i\partial_{y} \,.
\end{equation}
In order to perform the perturbative expansion, let us write the first three equations:
\begin{subequations}
  \label{eq:perturbative_exp_fk}
  \begin{align}
    \pderiv{\rho}{t} & =  -v_{0} {\rm Re}(\triangledown^* f_{1}) + \eta \nablab \cdot \left(\rho \nablab U \right)  \label{eq:perturbative_exp_fk_0} \\
    \pderiv{f_1}{t} & =  -\alpha f_1 - \frac{v_{0}}{2}
(\triangledown \rho + \triangledown^{*} f_2) + \eta \nablab \cdot \left(f_1 \nablab U \right) \label{eq:perturbative_exp_fk_1} \\
    \pderiv{f_2}{t} & =  -\alpha f_2 - \frac{v_{0}}{2}
( \triangledown f_1 + \triangledown^* f_3)  + \eta \nablab \cdot \left(f_2 \nablab U \right) . \label{eq:perturbative_exp_fk_2} 
  \end{align}
\end{subequations}
For $k\geqslant 1$, it is not difficult to see that $f_{k}$ has a fast relaxation dynamics for large $\alpha$, so that the time derivative $\partial_t f_k$ can be neglected on time scales large compared to $\alpha^{-1}$.
Under these assumptions, we obtain 
\begin{subequations}
  \label{eq:stationary_values_fk}
  \begin{align}
  f_1  & = -\frac{v_{0}}{2\alpha} \left(\triangledown \rho + \triangledown^* f_2 \right) + \frac{\eta}{\alpha} \nablab \cdot \left(f_1 \nablab U \right)   \label{eq:stationary_values_fk-1}
  \\
  f_2 & =  -\frac{v_{0}}{2\alpha} \left(\triangledown f_1 + \triangledown^* f_3 \right) + \frac{\eta}{\alpha} \nablab \cdot \left(f_2 \nablab U \right)  \, .  \label{eq:stationary_values_fk-2}
\end{align}
\end{subequations}
Fixing the large scale diffusion coefficient $D=v_{0}^{2}/2\alpha$, one obtains that $v_{0}\sim \alpha^{1/2}$. 
A careful look at Eq.~(\ref{eq:hierarchy_eqs_Fourier_RTP_2D}) indicates that a consistent scaling is given by $f_k \sim \alpha^{-|k|/2}$.
Then, by inspection, one sees from Eqs.~(\ref{eq:perturbative_exp_fk}) that to expand the continuity equation (\ref{eq:perturbative_exp_fk}a) to order $\alpha^{-1}$, one needs to expand $f_1$ to order $\alpha^{-3/2}$, and $f_2$ to order $\alpha^{-1}$.
At these orders, $f_{1}$ and $f_{2}$ read
\begin{subequations}
  \label{eq:f1_f2_RTP_2D}
  \begin{align}
    f_2 & = \frac{D}{2\alpha} \triangledown^2 \rho \label{eq:f1_f2_RTP_2D-2} \\
    f_1 & =  - \left[ \frac{D^{1/2}}{(2\alpha)^{1/2}} \triangledown \rho + \frac{D^{3/2}}{(2\alpha)^{3/2}}\Delta \triangledown \rho \right. \label{eq:f1_f2_RTP_2D-1}  \\
    &   \qquad\qquad\left. + \eta  \frac{(2D)^{1/2}}{2\alpha^{3/2}} \nablab \cdot\left(\triangledown \rho \nablab U\right) \right] , \nonumber
\end{align}
\end{subequations}
where $\Delta$ is the Laplacian.
One eventually gets a closed equation on the density field $\rho(\mathbf{r})$ which reads
\begin{align}
  \label{eq:closed_equation_RTP_2D}
   \pderiv{\rho}{t} = & \, D \nablab \cdot \left[ \nablab \left(\rho + \frac{D}{2\alpha}\Delta \rho \right) \right] \\
                       & + \frac{ \eta  D}{\alpha} \nablab \cdot \left[ \Delta \rho \nablab U + (\nablab \rho \cdot \nablab ) \nablab U \right] \notag \\
                       & + \eta  \nablab \cdot\left(\rho \nablab U \right) . \notag
\end{align}
Looking for a stationary perturbative solution, we propose the following ansatz
\begin{equation}
  \label{eq:ansatz_stationary_sol_RTP_2D}
  \rho(\bm{r}) \propto \exp\left( - \phi_{0}(\bm{r}) - \frac{1}{\alpha}\phi_{1}(\bm{r}) \right) \; ,
\end{equation}
where $\phi_{0}$ and $\phi_{1}$ do not depend on $\alpha$.
Eq.~(\ref{eq:closed_equation_RTP_2D}) can be rewritten in terms of the current ${\bf J}$ of particles, as $\partial_t \rho = \nablab \cdot {\bf J}$.
We now assume that $U(\bm{r})=U(x)$, i.e., that $U$ does not depend on $y$. In this case, the stationary current vanishes, and one finds
\begin{subequations}
  \label{eq:stationary_phi_RTP2D}
  \begin{align}
  \phi_{0}(x) & = \frac{\eta }{D} U(x) \label{eq:stationary_phi_RTP2D-0} \\
  \phi_{1}(x) & = - \frac{\eta}{2}U^{\prime\prime}(x) - \frac{\eta^2}{4D}\left(U^{\prime}(x)\right)^{2} + \frac{\eta^3}{2D^{2}} \int_{0}^{x}\ddr q \left(U^{\prime}(q)\right)^{3} . \label{eq:stationary_phi_RTP2D-1}
\end{align}
\end{subequations}
Note that $D/\eta$ plays the role of an effective temperature. At equilibrium, this ratio would correspond to the true thermodynamic temperature, from the fluctuation-dissipation theorem.
 
More importantly, one notices the presence of a non-local contribution of the external potential barrier at order $\alpha^{-1}$, because the term
$\int_{0}^{x}\ddr q \left(U^{\prime}(q)\right)^{3}$ does not vanish in the bulk, contrary to the other terms that involve only local derivatives of the potential.
This cubic contribution is quite similar to the 1D case at order $\alpha^{-1}$ and gives different contributions to $\Delta Q_{k}$ in \eqref{eq:delta_Q-correction_RTP_1D} depending on $k=A, \, B$. 
As a result, the stationary densities $\rho_A$ and $\rho_B$ depend on the detailed shape of the potential energy barrier.

\subsection{Active Brownian Particles in 2D}

For the active Brownian particles, the dynamics is very similar except for the angle that follows a Brownian motion rather than a jump process. The corresponding master equation reads 
\begin{align}
  \label{eq:master_eq_ABP_2D}
  & \pderiv{P}{t}(\bm{r}, \theta) = \\
  & \qquad - \nablab \cdot \big[ \big(v_{0} \bm{e}(\theta)
   - \eta \nablab U(\bm{r}) \big) P(\bm{r},\theta) \big] + D_{\rm r}\frac{\partial^{2} P}{\partial \theta^{2}} . \notag
\end{align}
Projecting this Fokker-Planck equation on the Fourier basis \eqref{eq:def:Fourier_transform_2D}, one obtains
\begin{align}
  \label{eq:projection_fourier_ABP_2D}
  \pderiv{f_{k}}{t} = & - \frac{v_{0}}{2}\left(\triangledown f_{k-1} + \triangledown^* f_{k+1} \right)  \\
  & + \eta  \nablab  \cdot \left(f_{k} \nablab U \right) - k^{2}D_{\rm r}  f_{k} . \notag
\end{align}
For ABPs, the quasi-equilibrium limit is obtained by taking $D_{\rm r}\to \infty$ while keeping $D=v_{0}^{2}/2D_{\rm r}$ fixed. It turns out that up to $k=2$, the hierarchy of equations is essentially the same as for RTP, up to the replacement of $\alpha$ by $D_{\rm r}$. The only difference is in the equation for $k=2$, where a factor $k^{2}=4$ appears in the angular diffusion term.
Taking into account the fast relaxation of the $f_{k}$ for $k \ge 1$, the closed equation over the density field $\rho(\bm{r})$ reads
\begin{align}
  \label{eq:closed_equation_ABP_2D}
  \pderiv{\rho}{t} = & \, D \nablab \cdot \left[ \nablab \left(\rho + \frac{D}{8D_{\rm r}}\Delta \rho \right) \right] \\
                      & + \frac{\eta  D}{D_{\rm r}} \nablab \cdot \left[ \Delta \rho \nablab U + (\nablab \rho \cdot \nablab) \nablab U \right] \notag \\
                      & + \eta  \nablab \cdot\left(\rho \nablab U \right). \notag
\end{align}
Eventually, looking for a stationary solution with a similar ansatz \eqref{eq:ansatz_stationary_sol_RTP_2D} as in the RTP case,
\begin{equation}
  \label{eq:ansatz_stationary_sol_ABP_2D}
  \rho(\bm{r}) \propto \exp\left( - \phi_{0}(\bm{r}) - \frac{1}{D_{\rm r}}\phi_{1}(\bm{r}) \right)
\end{equation}
one gets, assuming $U(\bm{r})=U(x)$,
\begin{subequations}
  \label{eq:stationary_phi_ABP2D}
  \begin{align}
  \phi_{0}(x) &= \frac{\eta}{D} U(x) \label{eq:stationary_phi_ABP2D-0} \\
 \phi_{1}(x) &= - \frac{\eta}{8} U^{\prime\prime}(x) - \frac{13\eta^2}{16 D}\left(U^{\prime}(x)\right)^{2} + \frac{7\eta^3}{8D^{2}} \int_{0}^{x}\ddr q \left(U^{\prime}(q)\right)^{3} \label{eq:stationary_phi_ABP2D-1}
\end{align}
\end{subequations}
Conclusions are the same as for RTPs in 2D since the expressions are very closed to each other, except for numerical prefactors. 
In particular, one recovers that the stationary densities $\rho_A$ and $\rho_B$ depend on the detailed shape of the potential energy barrier.

\begin{figure}[h]
  \centering
  \includegraphics[width=0.9\linewidth]{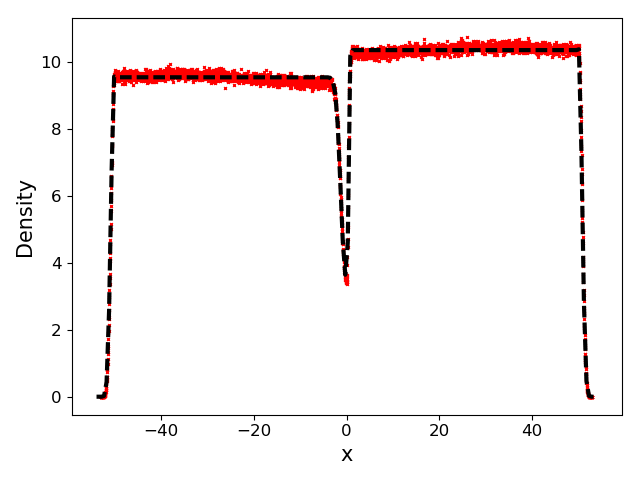}
  \caption{(Color online) Stationary density profile of ABPs for $D_{r}=30$: numerical simulations (red full line), perturbative analytical prediction (dashed line). The potential energy barrier approximately takes the form of two half-Gaussians: $U(x) = U_{0} \exp{-(\tfrac{x}{l(x)})^{2}}$ with $l(x)$ a smooth step function interpolating between $l_{A}=1$ and $l_{B}=0.3$. The confining walls at the edges are defined by quadratic potentials. Parameters: number of particles $N=1000$, linear sizes $L_x=L_y=100$, translational diffusion coefficient $D=\tfrac{v_{0}^{2}}{2D_{r}} = 1$, barrier height $U_{0}=1$, mobility $\eta = 1$.}
  \label{fig:density_profile_ABP}
\end{figure}

As a check of the validity of the perturbative expansion, Fig.~\ref{fig:density_profile_ABP} compares the stationary density profile (averaged in the $y$ direction) measured in a numerical simulation of ABPs for a reasonably large value of $D_{r}$, with the perturbative analytical prediction. The potential barrier is uniformly smooth but asymmetric. This asymmetry generally leads to different densities in the bulks of each subsystems ($x>0$ and $x<0$) contrary to the equilibrium situation (see details in subsection \ref{sec:chem_pot}).

\subsection{Active Brownian particles in an external field}

As a last example, we consider the case of ABP with a stochastic reorientation in a given direction $\theta_0$, with some noise. More precisely, with a rate $\lambda$, a new direction $\theta'$ is chosen randomly from a distribution $\psi(\theta'-\theta_0)$. For simplicity, the distribution $\psi(\theta)$ is assumed to be symmetric, $\psi(-\theta)=\psi(\theta)$.
As mentioned in the introduction, this model is inspired by the dynamics of the microalgae {\it Chlamydomonas Reinhardtii}, which can reorient its direction of motion towards a light source.\cite{Rafai2016}
On a theoretical level, it can be thought of as a way to include an external field in a model of active particles.
The evolution equation for the joint distribution $P(\bm{r}, \theta)$ reads
\begin{align}
  \label{eq:master_eq_reorient_2D}
  & \pderiv{P}{t}(\bm{r}, \theta) = \\
  & \qquad - \nablab \cdot \big[ \big(v_{0} \bm{e}(\theta)
   - \eta \nablab U(\bm{r}) \big) P(\bm{r},\theta) \big]  \notag \\
  &\qquad + D_{\rm r}\frac{\partial^{2} P}{\partial \theta^{2}} 
-\lambda P(\bm{r}, \theta) + \lambda \rho \psi(\theta-\theta_0) . \notag
\end{align}
Expanding in angular Fourier modes, one has
\begin{align}
  \label{eq:projection_fourier_reorient_2D}
  \pderiv{f_{k}}{t} = & - \frac{v_{0}}{2}\left(\triangledown f_{k-1} + \triangledown^* f_{k+1} \right)  \\
  & + \eta  \nablab  \cdot \left(f_{k} \nablab U \right) - k^{2}D_{\rm r}  f_{k} 
-\lambda f_k + \lambda \rho \psi_k e^{ik\theta_0} \notag
\end{align}
with $\psi_k = \int_{-\pi}^{\pi} d\theta \, \cos(k\theta) \psi(\theta)$.
Similarly to the case of ABP, we perform an expansion at large angular diffusion coefficient $D_r$, to leading order in $1/D_r$.
From now on, we assume that the potential energy is invariant along the direction $y$, and we choose $\theta_0=\frac{\pi}{2}$ so that the external field is parallel to the energy barrier.
The stationary equation for the density profile $\rho(x)$ (invariant in the $y$ direction) reads
\begin{equation}
\rho \eta U' + D_{\rm eff} \rho' + \frac{D^2}{8D_{\rm r}} \rho''' +\frac{\eta D}{D_{\rm r}}\frac{d}{dx}(U'\rho')=0
\end{equation}
with
\begin{equation}
D_{\rm eff} = D-\frac{\lambda D}{D_{\rm r}}\left(1+\frac{\psi_2}{4}\right)
\end{equation}
Assuming the form Eq.~\eqref{eq:ansatz_stationary_sol_ABP_2D} for the density $\rho(x)$, we eventually obtain
\begin{subequations}
  \label{eq:stationary_phi_reorient2D}
  \begin{align}
  \phi_{0}(x) &= \frac{\eta}{D} U(x)
\label{eq:stationary_phi_ABP2D-0field} \\
 \phi_{1}(x) &= \frac{\eta\lambda}{D} \left( 1+\frac{\psi_2}{4}\right)U(x) -\frac{\eta}{8}U^{\prime\prime}(x) \notag \\
 & \quad - \frac{13\eta^2}{16D}\left(U^{\prime}(x)\right)^{2} + \frac{7\eta^3}{8D^{2}} \int_0^{x}\ddr q \left(U^{\prime}(q)\right)^{3}  \; .
\label{eq:stationary_phi_reorient2D-1field}
\end{align}
\end{subequations}

Again, conclusions are the same as the two previous 2D independent active particles models since expressions are very similar, except for the non-crucial term proportional to $\lambda$ in $\phi_{1}$ that results from the presence of the external field.
As a result, we see that the dependence of the bulk densities $\rho_A$ and $\rho_B$ on the detailed shape of the potential energy barrier is a generic effect, that does not require any fine-tuning of parameters.

\subsection{Chemical potentials for independent active particles in 2D}
\label{sec:chem_pot}

In the three active particles models in 2D, the stationary density profile $\rho(x)$ has been computed perturbatively and generally reads
\begin{equation}
  \label{eq:stat_distrib_2D_actpart}
  \rho(x) \propto e^{-\phi(x)}
\end{equation}
with $ \phi(x)= \phi_{0}(x) + \tfrac{1}{\zeta}\phi_{1}(x)$, $\zeta = \alpha \text{ or } D_{\rm r}$ depending on the model considered.

Since particles are independent, the dynamics on the global density in each subsystems A and B is also expected to be Poissonian as for the RTP in 1D. Therefore, the large deviation function $\mathcal{I}(\rho_{A},\rho_{B})$ associated with $P(\rho_{A}|\bar{\rho})$ (see Eq.~\eqref{eq:large_dev_proba_density_A_RTP_1D}) is exactly the same as eq.~\eqref{eq:large_dev_proba_density_A_RTP_1D_I}.
The stationary densities in the bulks of each systems, $\rho_{A}^{\ast}$ and $\rho_{B}^{\ast}$, are obtained from the condition that the density profile is continuous in $x=0$ (a condition valid here because we have considered an equal speed $v_0$ in both compartments), leading to the condition
\begin{equation}
  \label{eq:rhoA-rhoB_2D_act_part}
  \rho_{A}^{\ast} e^{\phi(x_{A}^{\ast})} = \rho_{B}^{\ast} e^{\phi(x_{B}^{\ast})} \; .
\end{equation}
with $x_{k}^{\ast}$, $k=A,\, B$, two arbitrary points in the uniform bulks of systems $A$ and $B$. Contrary to the equilibrium situation ($\zeta \to \infty$) for which $\phi(x)=\beta U(x)$, the presence of a non-local contribution proportional to $\int_{0}^{x}\mathrm{d}q \, [U'(q)]^{3}$ in $\phi(x)$ for each model leads to an additional bias as long as $\zeta^{-1} >0$.

\begin{figure}[h]
  \centering
  \includegraphics[width=0.9\linewidth]{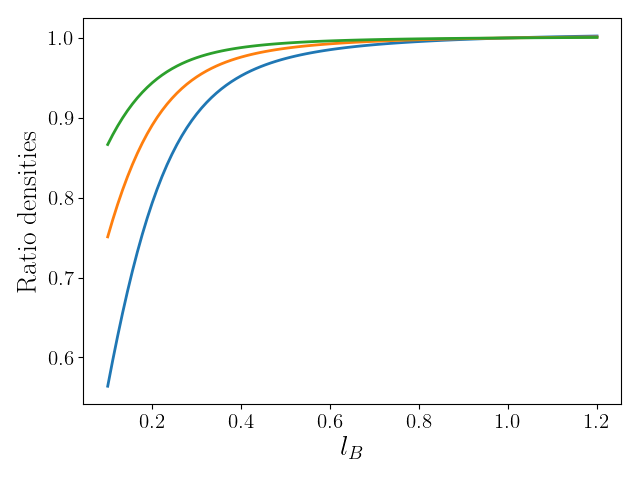}
  \caption{(Color online) Ratio of bulk densities $\tfrac{\rho_{A}^{\ast}}{\rho_{B}^{\ast}}$ for ABPs, with respect to the asymmetry parameter of the potential $l_{B}$, for three values of $\zeta = D_{\rm r} = 25$ (blue), $50$ (orange), $100$ (green). The potential barrier is approximately given by two half-Gaussians: $U(x) = U_{0} \exp{-(\tfrac{x}{l(x)})^{2}}$, where $l(x)$ is a smooth step function interpolating between $l_{A}=1$ (fixed) and $l_{B}$ (varied).
Densities $\rho_{A}^{\ast}$ and $\rho_{B}^{\ast}$ equalize the chemical potentials $\mu_A^{\rm cont}$ and $\mu_B^{\rm cont}$ defined in Eq.~(\ref{eq:mu_actpart_2D}).
For ABPs, $K=\tfrac{7}{8}$ in Eq.~(\ref{eq:delta_Qk_actpart_2D}).
}
  \label{fig:ratio_dens}
\end{figure}

The associated chemical potentials are thus very similar to the RTP in 1D \eqref{eq:mu_cont_RTP_1D} and read
\begin{equation}
  \label{eq:mu_actpart_2D}
  \mu_{k}^{\rm cont}(\rho_{k}) = \ln \rho_{k} - \frac{1}{\zeta}\Delta Q_{k} \, ,
\end{equation}
with
\begin{equation}
  \label{eq:delta_Qk_actpart_2D}
  \Delta Q_{k} = K \frac{\eta^{3}}{D^{2}} \int_{x_{k}^{\ast}}^{0}\mathrm{d}q \, [U'(q)]^{3} \, ,
\end{equation}
$K$ being a numerical constant which is equal to $\tfrac{1}{2}$ for RTP and $\tfrac{7}{8}$ for both ABP and ABP with an external field.
Equalization of the chemical potentials $\mu_A^{\rm cont}$ and $\mu_B^{\rm cont}$ yields back the densities $\rho_{A}^{\ast}$ and $\rho_{B}^{\ast}$.
Fig.~\ref{fig:ratio_dens} displays the ratio $\rho_{A}^{\ast}/\rho_{B}^{\ast}$ as a function of the asymmetry of the potential barrier.
A strong effect is observed, even for relatively large values of $D_{\rm r}$.

We thus conclude that the nonequilibrium chemical potential defined in models of independent active particles does not satisfy an equation of state in general, as it does not depend only on the bulk density, but also on the shape of the energy barrier.
As mentioned in Ref.~\onlinecite{Guioth2018}, the zeroth law of thermodynamics can be restored if one includes the (half) barrier in the definition of the system.

\section{Relation between chemical potential and pressure}

The lack of an equation of state for the chemical potential is reminiscent of
a similar situation occuring for the pressure in models of active particles. As mentioned in the introduction, many studies on the notion of pressure in active systems have been conducted recently \cite{solon2015prl,solon2015nat,solon2018generalized-njp,fily2017mechanical,solon2018generalized-pre,Joyeux2016}.
The mechanical pressure can be computed as the average force exerted by the particles present in compartment $k=A$, $B$ on the wall separating the two compartments:
\begin{equation}
  \label{eq:def:pressure_RTP}
  P_{k \to {\rm wall}}(\rho_{k}^{\ast}) = \int_{x_{k}^{\ast}}^{0} \rho(x) U'(x) \, \ddr x \, .
\end{equation}
(we recall that $x_{k}^{\ast}$ is a point in the bulk of compartment $k$).
When studying the pressure, the potential energy barrier is assumed to be infinite. In constrast, in the study of the chemical potential, the barrier needs to be finite (though high to implement a weak contact) to allow for particle transfer between compartments.

For simplicity, we discuss here the connection between chemical potential and pressure in the framework of the one-dimensional RTP model, but generalization to other models is straightforward.
If the system behaves as if it was at equilibrium (which can be achieved in the 1D RTP by taking the limit $\alpha\to\infty$, $D=v_{0}^{2}/\alpha$ kept fixed with $\beta = \eta /D$ the effective temperature), the density profile reads
\begin{equation}
\rho_k(x) = \rho_k^{\ast} \,e^{-\beta U(x)}\,,
\end{equation}
(where $\rho_k^{\ast}$ is the bulk pressure in compartment $k$),
leading to a pressure on the side $k$ of the wall
\begin{equation}
P_{k\to {\rm wall}} = \beta^{-1} \rho_{k}^{\ast} \,,
\end{equation}
consistently with what one would expect for a perfect gas.
But since $\rho(x)$ is non-local in $U(x)$ for RTPs in 1D,
$P_{k\to {\rm wall}}$ has been shown to depend on the external potential barrier profile $U(x)$. It turns out that the pressure can be exactly computed here (see, e.g., appendix of Ref.~\onlinecite{solon2015nat} in the case of an infinite barrier):
\begin{equation}
  \label{eq:pressure_RTP1D_exact_computation}
  P_{k\to {\rm wall}}(\rho_{k}^{\ast}) = \frac{\rho_{k}^{\ast}v_{k}^{2}}{\eta \alpha} \left(1 - e^{-\Delta Q_{k}}\right) \; .
\end{equation}

At equilibrium, there is a well-known relationship between pressure and chemical potentials which is often called a Maxwell relation (see Refs.~\onlinecite{sekimoto2010stochastic,sasa2006steady} or Refs.~\onlinecite{takatori2015towards,takatori2014swim} in the context of active matter). The Maxwell relation reads
\begin{equation}
  \label{eq:maxwell_relation_RTP}
  \pderiv{P}{\rho} = \rho \pderiv{\tilde{\mu}}{\rho} \; .
\end{equation}
It is important to note that the chemical potential $\tilde{\mu}$ appearing in the Maxwell relation (\ref{eq:maxwell_relation_RTP}) has the dimension of an energy, and is related to the dimensionless chemical potential $\mu$ considered in this paper through $\mu=\beta\tilde{\mu}$, with $\beta$ the inverse temperature of the equilibrium system considered.
Out of equilibrium, the temperature of the system is, strictly speaking, ill-defined. Yet, it is possible to circumvent the problem by defining an effective temperature $\beta_{\rm eff}^{-1}=v_{0}^{2}/(\alpha \eta)$, and a chemical potential
$\tilde{\mu}=\beta_{\rm eff}^{-1}\mu$.
In this case, one can wonder whether the Maxwell relation is still valid.
For independent RTPs in one dimension, one obtains
\begin{equation}
  \label{eq:violation_maxwell_eq_RTP1D:mu}
  \rho_{k}\pderiv{\tilde{\mu}_{k}}{\rho_{k}} =  \frac{v_{k}^{2}}{\alpha \eta }
\end{equation}
and
\begin{equation}
  \label{eq:violation_maxwell_eq_RTP1D:p}
 \pderiv{P_{k\to {\rm wall}}}{\rho_{k}} = \frac{v_{k}^{2}}{\alpha \eta } \left( 1 - e^{-\Delta Q_{k}} \right) \,.
\end{equation}
Hence strictly speaking, the Maxwell relation is not valid.
We note, however, that in the weak contact limit needed to define the nonequilibrium chemical potential\cite{Guioth2018}, the energy barrier separating compartments A and B is high, implying that $\Delta Q_{k}$ defined in Eq.~(\ref{eq:delta_Q-correction_RTP_1D}) is large. It thus follows that the correction to the Maxwell equation is small in this regime. In addition, it might also be argued that the definition of the effective temperature $\beta_{\rm eff}^{-1}$ is phenomenological and not necessarily well grounded, so that including appropriate corrections to this effective temperature might restore the Maxwell relation (on condition, of course, that proper justifications to the modified effective temperature be given).

It turns out that more general arguments in favor of the breaking of the Maxwell relation can be given. The first argument goes as follows.
As shown in Ref.~\onlinecite{sasa2006steady}, the Maxwell relation is intimately related to a balance between forces applied by an external potential on the particles and by the pressure exerted by the particles. Yet, since the two stationary densities $\rho_{k}^{\ast}$ for $k=A$, $B$ are related through the equality of currents \eqref{eq:equalization_current_RTP_1D}, one finds that
\begin{equation}
  \label{eq:ratio_pressure_RTP1D}
  \left| \frac{P_{B \to {\rm wall}}}{P_{A\to {\rm wall}}} \right| = \frac{v_{B}}{v_{A}}\, \frac{e^{\Delta Q_{B}} - 1}{e^{\Delta Q_{A}} - 1} \neq 1 \; ,
\end{equation}
as long as $v_{A}\neq v_{B}$ and $\Delta Q_{A} \neq \Delta Q_{B}$. Hence, there exists a net force applied on the wall\cite{solon2015nat}, which needs to be balanced to keep the potential energy barrier $U(x)$ centered at $x=0$.
In reaction, the wall exerts a net force on the particles which needs in turn to be balanced for the average current to vanish. Such a balance is provided by the active force, understood as the force exerted by the environnement on the active particles.\cite{fily2017mechanical} This means that at the end, one should experimentally observe a deformation or a flow in the substrate.
As a consequence, if the force on the wall is not balanced, the wall is moving\cite{solon2015nat,Joyeux2016}.

%Eventually, no link seems to emerge between the mechanical pressure \eqref{eq:def:pressure_RTP} and the chemical potentials at contact. This is expected to be so because of the lack of balance between the forces exerted by the external potential and the pressure of the particles, but further studies in this direction should definitely be performed to validate more systematically this result.

An alternative interpretation of the breaking of the Maxwell relation comes from the large deviation approach, which lies at the root of our definition of the chemical potential. 
At equilibrium, the pressure $P$ and the chemical potential $\tilde{\mu}$ are directly related to the free energy $F$ through the definitions
$P=-\partial F/\partial V$ and $\tilde{\mu}=\partial F/\partial V$.
The scaling relation $F(N,V)=V\, f(N/V)$ straightforwardly leads to the Maxwell relation \eqref{eq:maxwell_relation_RTP}.
In terms of large deviation functions, this means that the large deviation function associated with an exchange of particles is $I_N(\rho)=f(\rho)$, and
the one associated with an exchange of volume is $I_V(v)=v f(1/v)$, with the same function $f$ in both cases.
Following Ref.~\onlinecite{Bertin06,Bertin07,Guioth2018}, the definition of an intensive parameter conjugated to a conserved quantity can be extended out of equilibrium through the use of large deviation functions.
However, as shown in Ref.~\onlinecite{Guioth2018}, the large deviation function generically depends on the contact dynamics, so that exchanging particles or volume no longer leads to the same scaling function $f$, at odds with equilibrium
(the thermodynamic pressure could even be ill-defined\cite{Guioth2018}).
%Actually, the large deviation function in the case of volume exchange may not even satisfy the additivity property required for the definition of an intensive thermodynamic parameter \cite{Bertin06,Bertin07,Guioth2018}.
This difference with the equilibrium case is thus another reason explaining the breaking of the Maxwell relation \eqref{eq:maxwell_relation_RTP}.

\section{Conclusion}

In this paper, we have discussed the notion of nonequilibrium chemical potential in the simple framework of gases of noninteracting active particles placed in two compartments separated by a high energy barrier. %---a way to implement in practice the notion of weak contact \cite{sasa2006steady,Guioth2018}.
Our definition of a nonequilibrium chemical potential is well-grounded in the theory of large deviation functions\cite{Guioth2018}, and does not rely on the assumed validity of the Maxwell relation connecting pressure and chemical potential\cite{takatori2015towards}. The main prediction of Ref.~\onlinecite{Guioth2018} is that the nonequilibrium chemical potential generically depends on the contact dynamics. We have confirmed this result for several models of noninteracting active particles in one and two dimensions.

Although for two-dimensional systems our arguments are based on a perturbative expansion in the vicinity of an effective equilibrium limit, this result is enough to show qualitatively that the nonequilibrium chemical potential generically depends on the detailed shape of the potential energy barrier between the two compartments. In other words, the nonequilibrium chemical potential does not satisfy an equation of state in terms of bulk properties of the gas of active particles, but the shape of the potential barrier has to be taken into account.
As already mentioned above, this result is reminiscent of a similar result obtained when evaluating the mechanical pressure exerted by active particles on a (soft) wall modeled by a diverging potential energy profile.
An interesting extension of this work could be to try to incorporate interactions between particles in the evaluation of the chemical potential.
Although no conclusion can be drawn at this stage, it is unlikely that the inclusion of interactions could restore a bulk equation of state. At least this is not the case when considering pressure \cite{solon2015nat}.

%%%%%%%%%%%%%%%%%%%%%%%%%%%%%%%%%%%%%%%%%%%%%%%%%%%%%%%%% 
%%%%%%%%%%%%%%%%%%%%%%%%%%%%%%%%%%%%%%%%%%%%%%%%%%%%%%%%%

%%% BIBLIOGRAPHY %%%%%%%%%%%%%%%%%%%%
\bibliography{biblio_contact_actpart}
%%%%%%%%%%%%%%%%%%%%%%%%%%%%%%%%%%%%%

\end{document}